\documentclass[a4paper]{JHEP3}

\usepackage{epsfig,multicol,bbm}
\usepackage{amsmath}
\usepackage{amssymb}
\usepackage{amsfonts}
\usepackage{graphicx,epsfig}
\usepackage{subfigure}
\usepackage{exscale}
\usepackage{float}
\usepackage[numbers,sort&compress]{natbib}

\newcommand\fverb{\setbox\fverbbox=\hbox\bgroup\verb}
\newcommand\fverbdo{\egroup\medskip\noindent%
			\fbox{\unhbox\fverbbox}\ }
\newcommand\fverbit{\egroup\item[\fbox{\unhbox\fverbbox}]}
\newbox\fverbbox

\newcommand{\Z}{{\mathbb{Z}}}

\newcommand{\p}{\partial}

\title{Linear broadening of the confining string in Yang-Mills theory at
low temperature}

\author{F.\ Gliozzi\\
Dipartimento di Fisica Teorica, Universit\`a di Torino, and \\
INFN, Sezione di Torino, via P.\ Giuria 1, 10125 Torino, Italy. \\
	E-mail: \email{gliozzi@to.infn.it}}

\author{M.\ Pepe\\
INFN, Sezione di Milano-Bicocca\\ 
Edificio U2, Piazza della Scienza 3, 20126 Milano, Italy. \\
	E-mail: \email{pepe@mib.infn.it}}

\author{U.-J.\ Wiese\\
Albert Einstein Center for Fundamental Physics \\
Institute for Theoretical Physics, Bern University\\
Sidlerstr.\ 5, 3012 Bern, Switzerland.\\
	E-mail: \email{wiese@itp.unibe.ch}}

\abstract{The logarithmic broadening predicted by the systematic low-energy effective 
field theory for the confining string has recently been verified in numerical 
simulations of $(2+1)$-d $SU(2)$ lattice Yang-Mills theory at zero temperature. 
The same effective theory predicts linear broadening of the string at low 
non-zero temperature.
In this paper, we verify this prediction by comparison with very 
precise Monte Carlo data. The comparison involves no additional adjustable 
parameters, because the low-energy constants of the effective theory have 
already been fixed at zero temperature. It yields very good agreement between
the underlying Yang-Mills theory and the effective string theory.}


\begin{document} 

\section{Introduction}

At low temperature, in Yang-Mills theory static quarks and anti-quarks are 
confined to one another by an unbreakable string. The string itself is an
interesting object with peculiar dynamical properties. The relevant energy scale
for these dynamics is set by the string tension $\sigma$, which defines the 
asymptotic slope of the linearly rising quark-anti-quark potential. The string 
tension also controls the amplitude of transverse fluctuations of the string. 
As was pointed out a long time ago by L\"uscher, Symanzik, and Weisz 
\cite{Lue80}, these fluctuations are described by a systematic low-energy 
effective string theory. In fact, the transverse fluctuations are massless 
modes (i.e.\ Goldstone excitations) of a spontaneously broken translational 
symmetry of the string world-sheet, also known as capillary waves in condensed
matter physics. The effective string theory plays the role of chiral 
perturbation theory for these Goldstone modes. Using the effective theory, 
L\"uscher has calculated the leading correction to the linear quark-anti-quark 
potential
\begin{equation}
\label{potential}
V(r) = \sigma r - \frac{\pi (d-2)}{24 r} + {\cal O}(1/r^3),
\end{equation}
where $d$ is the space-time dimension and $(d-2)$ counts the directions
transverse to the string world-sheet \cite{Lue80,Lue81}. 
Accurate numerical simulations have shown the validity
of that expectation \cite{has,Cas96a,Jug02,Lue02,Cas04,Cas06,Har06,Bri08,Ath08,Bra09}.
The effective theory also accounts for the 
string width \cite{Lue81a}. Due to its transverse fluctuations, the string 
broadens as the quark sources are separated. The cross-sectional area swept out 
by the fluctuating string increases logarithmically with the distance $r$ 
between the static sources, i.e.
\begin{equation}
\label{width}
w^2_{lo}(r/2) = \frac{d-2}{2 \pi \sigma} \log(r/r_0).
\end{equation}
Here $w_{lo}(r/2)$ is the leading-order string width at the distance $r/2$,
half-way between the external static quark sources, and $r_0$ is a length scale 
that enters the effective string theory as a low-energy parameter. The above
expression for the width is universal and applies to strings in Yang-Mills
theory as well as to fluctuating interfaces between different phases of 
condensed matter. It has been verified in numerical simulations for the 
interfaces separating the two low-temperature phases of the 3-d Ising model
above the roughening transition \cite{cas96}. Similar results have been 
obtained in a $(2+1)$-d $\Z(4)$ gauge theory \cite{giu07}. Since numerical
lattice calculations of the string width in non-Abelian gauge theories are
computationally very challenging, early calculations were affected by large 
statistical errors and did thus not yield conclusive results \cite{bali95}.
Only recently, using the very efficient L\"uscher-Weisz multi-level algorithm 
\cite{Lue01,Lue02}, the logarithmically divergent string width has been 
accurately verified by simulating $(2+1)$-d $SU(2)$ lattice Yang-Mills theory 
\cite{Gli10}. Recent results on the effective string description of the color flux tube in 
$(2+1)$-d $SU(N)$ Yang-Mills theory can be found in \cite{Ath07,Bia09}.

At low temperature $T = 1/\beta$, the extent $\beta$ of the periodic Euclidean
time direction enters as another length scale in the low-energy effective 
string theory. At low $T \ll \sqrt{\sigma}$, the temperature effects on the 
string tension and the string broadening are again accessible to a systematic 
treatment. At leading order, the temperature-dependent string tension is given
by 
\begin{equation}
\sigma(\beta) = \sigma -\frac{\pi (d-2)}{6\beta^2}.
\end{equation}
Similarly, the leading-order string width at a distance $r/2 \gg \beta$ takes 
the form \cite{All08,Cas10}
\begin{equation}
w_{lo}^2(r/2) = \frac{d-2}{2 \pi \sigma} \log\frac{\beta}{4 r_0} +
\frac{d-2}{4\beta \sigma}\, r +{\cal{O}}(\mbox{e}^{-2\pi r/\beta}).
\label{linear}
\end{equation}
Hence, at non-zero temperature the confining string broadens linearly with the distance
between the external static quarks. Numerical evidence for linear string broadening at
finite temperature was presented for fluctuating interfaces in the 3-d Ising model in
\cite{All08,Cas10} and, recently, for $(3+1)$-d $SU(3)$ Yang-Mills theory in \cite{Bak10}.
In this paper, using the L\"uscher-Weisz multi-level simulation technique, we perform very
accurate Monte Carlo calculations of the temperature-dependent string width in $(2+1)$-d
$SU(2)$ Yang-Mills theory. The numerical results are then compared with the corresponding
analytic predictions of the systematic low-energy effective string theory at the 2-loop
level \cite{Gli10b}. Since the low-energy parameters of the effective theory were
previously fixed by a comparison with Monte Carlo data at zero temperature, there are no
adjustable parameters. We find perfect agreement between the numerical data and the
analytic predictions, which lends further support to the quantitative validity of the
effective theory.

The rest of this paper is organized as follows. Section 2 describes the analytic
calculations in the low-energy effective string theory, which are then compared 
with the numerical results obtained in section 3. Section 4 contains our 
conclusions.

\section{Analytic Results of the Low-Energy Effective String Theory}

Let us consider the systematic low-energy effective theory describing the 
transverse fluctuations of the string world sheet in $(2+1)$-d Yang-Mills 
theory. In terms of the height variable $h(x,t)$ the corresponding Euclidean
action is given by
\begin{equation}
\label{action3d}
S[h] = \int_0^\beta dt \int_0^r dx \ \frac{\sigma}{2} \left[\p_\mu h \p_\mu h -
\frac{1}{4} \left(\p_\mu h \p_\mu h\right)^2 \right].
\end{equation}
Here $x \in [0,r]$ and $t \in [0,\beta]$ parameterize the 2-d base-space. Since
the string ends at static quark charges separated by a distance $r$, its 
fluctuation field obeys the boundary condition $h(0,t) = h(r,t) = 0$.
The boundary conditions explicitly break translation invariance and one might
have expected boundary terms to be present in the effective action as well 
\cite{Lue02}. Remarkably, due to open-closed string duality, such terms are 
absent and the prefactor of the first sub-leading term is uniquely determined 
\cite{Lue04}. This behavior even extends to the next order six-derivative term.
Indeed, up to that order the effective action coincides with the one of the 
Nambu-Goto string \cite{aha09,aha10A,aha10B}.

The confining string of a lattice Yang-Mills theory suffers from lattice 
artifacts which disappear only in the continuum limit. Away from the continuum 
limit, in the strong coupling regime, the lattice string is rigid, i.e.\ it 
follows the discrete lattice steps and does not even support massless 
fluctuations. When one crosses the roughening transition at intermediate values 
of the coupling, one enters the rough phase with massless Goldstone excitations 
of the string, to which the low-energy effective string theory applies. Since 
the lattice theory is invariant only under discrete rotations and not under the 
full Poincar\'e group, before one reaches the continuum limit two additional 
terms proportional to $\sum_{\mu = 1,2} \p_\mu \p_\mu h \p_\mu \p_\mu h$ and to
$\sum_{\mu = 1,2} (\p_\mu h)^4$ enter the effective theory in the bulk. Since these terms 
contain four derivatives, they are of sub-leading order. Hence, they have no 
effect on the L\"uscher term or on the leading behavior of the string width. As 
a result, the L\"uscher term is indeed completely universal. Provided its 
world-sheet is rough, even a lattice string supports exactly massless modes 
which contribute $- \pi/24 r$ to the static quark potential. In particular, 
this term is not affected by lattice spacing artifacts, since such effects are 
of sub-leading order. In the following, we will work sufficiently close to the 
continuum limit of the lattice theory. In that case, the effective action of 
eq.(\ref{action3d}) without the additional non-rotational-invariant terms is 
sufficient.

Recently, an analytic expression for the string width at zero temperature has 
been derived from the effective string theory at the 2-loop level \cite{Gli10b}
\begin{equation}\label{w2C}
w^2(r/2)=\left( 1 +\frac{4\pi f(\tau)}{\sigma r^2}\right) w^2_{lo}
(r/2)-\frac{f(\tau)+g(\tau)}{\sigma^2r^2},
\end{equation}
where
\begin{equation}
f(\tau)=\frac{E_2(\tau)-4E_2(2\tau)}{48}\, , \quad
g(\tau)=i \pi \tau \left(\frac{E_2(\tau)}{12}- q\frac {d}{dq} \right) 
\left( f(\tau)+\frac{E_2(\tau)}{16} \right)+\frac{E_2(\tau)}{96}.
\end{equation}
The parameter $q=\mbox{e}^{2\pi i \tau}$ depends on the ratio $\tau = i\beta/(2r)$ and 
\begin{equation}
E_2(\tau) = 1 - 24 \sum_{n=1}^\infty \frac{n q^n}{1-q^n}~,
\label{e2}
\end{equation}
is the first Eisenstein series. The leading-order expression for the width
\begin{equation}
w^2_{lo}(r/2) = \frac{1}{2 \pi \sigma} \log\left(\frac{r}{r_0}\right) + 
\frac{1}{\pi \sigma} \log\left(\frac{\eta(2 \tau)}{\eta(\tau)^2}\right),
\end{equation}
contains the Dedekind $\eta$-function
\begin{equation}
\eta(\tau) = q^{\frac1{24}} \prod_{n=1}^\infty(1-q^n).
\end{equation}
This function and the Eisenstein series $E_2$ obey the following transformations
under the inversion $\tau \to - \tau^{-1}$ 
\begin{equation}
\eta(\tau) = \frac{1}{\sqrt{-i\tau}} \eta(- \frac{1}{\tau})~, \quad
E_2(\tau) = \frac{1}{\tau^2} \, E_2(-\frac{1}{\tau} ) - \frac{6}{i \pi \tau}~.
\end{equation}
Using these relations, the expression of eq.(\ref{w2C}) in the zero-temperature
limit, $\beta \gg r $, is readily converted into an expression in the limit $r \gg
\beta$. In this case, the  temperature is still small, but the string length $r$ is much
larger than the inverse temperature $\beta$. 

For completeness, we present also the expression for the force $F$ at finite temperature,
defined as the first derivative of the potential. At leading order the force is given by
\begin{equation}
F_{lo}(r) = \sigma + \frac{1}{2 \beta r}-\frac{\pi}{6 \beta^2}\, E_2(-\frac{1}{\tau}).
\label{forceLO}
\end{equation}
The potential at next-to-leading order has been obtained from the computation of the
partition function using the effective string action of eq.(\ref{action3d}) \cite{Die82}. 
Then the force is given by the expression
\begin{equation}
F(r) = F_{lo}(r) -\frac{\pi^2}{72 \sigma \beta^4}\frac{d}{dr} 
\left[ r \left( 2 E_4(-\frac{1}{\tau})-E_2^2(-\frac{1}{\tau})\right) 
-\frac{9\beta^2}{\pi^2 r} +\frac{6 \beta}{\pi} E_2(-\frac{1}{\tau}) \right]
\label{forceNLO}
\end{equation}
This equation shows that the 2-loop string tension at finite temperature is given by
\begin{equation}
\sigma(\beta) = \sigma -\frac{\pi}{6 \beta ^2} -\frac{\pi^2}{72 \sigma \beta^4}.
\label{sigmaNLO}
\end{equation}

\section{Numerical Results for $(2+1)$-d $SU(2)$ Lattice Yang-Mills Theory}

Since it is least problematical in numerical simulations, we concentrate on
$(2+1)$-d $SU(2)$ Yang-Mills theory on a cubic lattice of size 
$L \times L \times \beta$, with the Euclidean time extent $\beta$ determining 
the inverse temperature. The partition function is then given by
\begin{equation}
Z = \int {\cal D}U \ \exp(- S[U]) = \prod_{x,\mu} \int_{SU(2)} dU_{x,\mu} \
\exp(- S[U]).
\end{equation}
Here $dU_{x,\mu}$ denotes the local gauge invariant Haar measure for a parallel 
transporter variable $U_{x,\mu} \in SU(2)$ located on the link $(x,\mu)$. We use
the standard Wilson plaquette action 
\begin{equation}
S[U] = - \frac{2}{g^2} \sum_{x,\mu,\nu} 
\mbox{Tr} [U_{x,\mu} U_{x+\hat\mu,\nu} U_{x+\hat\nu,\mu}^\dagger U_{x,\nu}^\dagger],
\end{equation}
where $g$ is the bare gauge coupling and $\hat\mu$ is a unit-vector pointing in 
the $\mu$-direction. As usual in lattice calculations, all physical quantities 
are expressed in units of the lattice spacing which is set to 1. 

The external static quarks at the two ends of the confining string are
represented by Polyakov loops
\begin{equation}
\Phi_x = \mbox{Tr} \left[\prod_{t=1}^\beta U_{x+t\hat 2,2}\right],
\end{equation}
which wind around the periodic Euclidean time direction. We choose the 
2-direction to represent Euclidean time. When two external quarks are separated
in the 1-direction, the world-sheet of the confining string extends in the 
space-time directions $\mu = 1,2$, while the string fluctuates in the 
transverse 3-direction. The static quark potential $V(r)$ is extracted from the 
Polyakov loop correlation function
\begin{equation}
\langle \Phi_0 \Phi_r \rangle = \frac{1}{Z} \int {\cal D}U \ \Phi_0 \Phi_r
\exp(- S[U]) \sim \exp(- \beta V(r)),
\end{equation}
in the zero-temperature limit $\beta \rightarrow \infty$. 

The force between the two static quarks is extracted from the discrete derivative of the
static potential
\begin{equation}
F(\bar r)= V(r+1)-V(r) = -\frac{1}{\beta}\, \log 
\frac{\langle \Phi_0 \Phi_{r+1} \rangle}{\langle \Phi_0 \Phi_r \rangle},
\end{equation}
where, writing explicitly the lattice spacing $a$, 
$\bar r = r + \frac{a}{2} + {\cal{O}}(a^2)$ is an improved definition of the point where
the force is computed \cite{Som93}. 

The width of the string is determined from the connected correlation function 
\begin{equation}\label{obs}
C(x_3) = \frac{\langle \Phi_0 \Phi_r P_x \rangle}{\langle \Phi_0 \Phi_r \rangle}
- \langle P_x \rangle,
\end{equation}
of a pair of Polyakov loops with a single plaquette 
\begin{equation}
P_x = \mbox{Tr} [U_{x,1} U_{x+\hat 1,2} U_{x+\hat 2,1}^\dagger U_{x,2}^\dagger],
\end{equation}
in the $1$-$2$ plane, which measures the color-electric field along the 
1-direction of the string as a function of the transverse displacement $x_3$.
The plaquette is located at the site $x = (r/2,t,x_3)$ and thus measures the 
transverse fluctuations of the color flux tube at the maximal distance $r/2$ 
from the external quark charges. 

We have performed numerical simulations at the bare gauge coupling $4/g^2 = 9.0$ on a
$90^2\times16$ lattice, corresponding to $T/T_c \approx 0.38$. We have used the
L\"uscher-Weisz technique \cite{Lue01,Lue02} with a single level and with slices of thickness 4.
In table \ref{upstep} we report the number of sub-lattice updates for 
various ranges of distances between the static sources.

\begin{table}[h]
\centering
\begin{tabular}{|c|c|}
\hline
$r$ range & updates \\
\hline
4-8   & 3200  \\
9-15  & 24000 \\
16-22 & 32000 \\
23-25 & 80000 \\
\hline
\end{tabular}
\caption{Number of sub-lattice updates of the single level L\"uscher-Weisz algorithm for
various ranges of distances between the static sources.}\label{upstep}
\end{table} 

\begin{figure}[htb]
\centering
\includegraphics[height=.6\textwidth,width=0.9\textwidth]{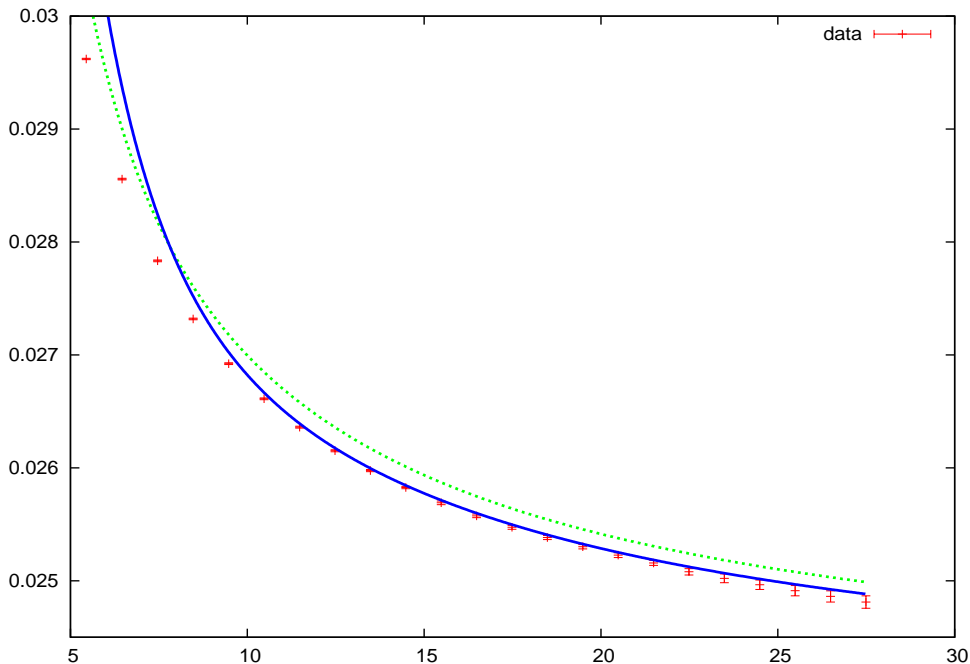}
\caption{The force $F(\bar r)$ as a function of the distance between the two static
quarks. The dotted line and the solid line are the leading order approximation of
eq.(\protect\ref{forceLO}) and the next-to-leading order approximation of
eq.(\protect\ref{forceNLO}), respectively.}
\label{force}\end{figure}

In figure \ref{force}, we plot the force between the two static quarks as a function of
the distance between the sources. The dotted line is the force in the leading order
approximation of eq.(\ref{forceLO}). The solid line is the next-to-leading order approximation
given by eq.(\ref{forceNLO}). It is important to note that there is no free parameter and
$\sigma = 0.025897(15)$ is the zero-temperature string tension. The next-to-leading order
expression already provides a very good description of the numerical data. Considering
the accuracy of the data, it is reasonable to expect that the residual tiny discrepancy
can be described by the next-to-next-to-leading order correction.

For the measurement of the string width, the correlation function is fit to the ansatz
\begin{equation}\label{fitbell}
\frac{\langle \Phi_0 \Phi_r P_x \rangle}{\langle \Phi_0 \Phi_r \rangle} = A\, \exp(- x_3^2/s)\; 
\frac{1 + B\, \exp(- x_3^2/s)}{1 + D\, \exp(- x_3^2/s)} + K,
\end{equation}
and the squared width of the string is then obtained as the second moment of 
the correlation function
\begin{equation}
w^2(r/2) = \frac{\int dx_3 \ x_3^2\, C(x_3)}{\int dx_3 \ C(x_3)}.
\end{equation}

\begin{figure}[htb]
\centering
\includegraphics[height=.6\textwidth,width=0.9\textwidth]{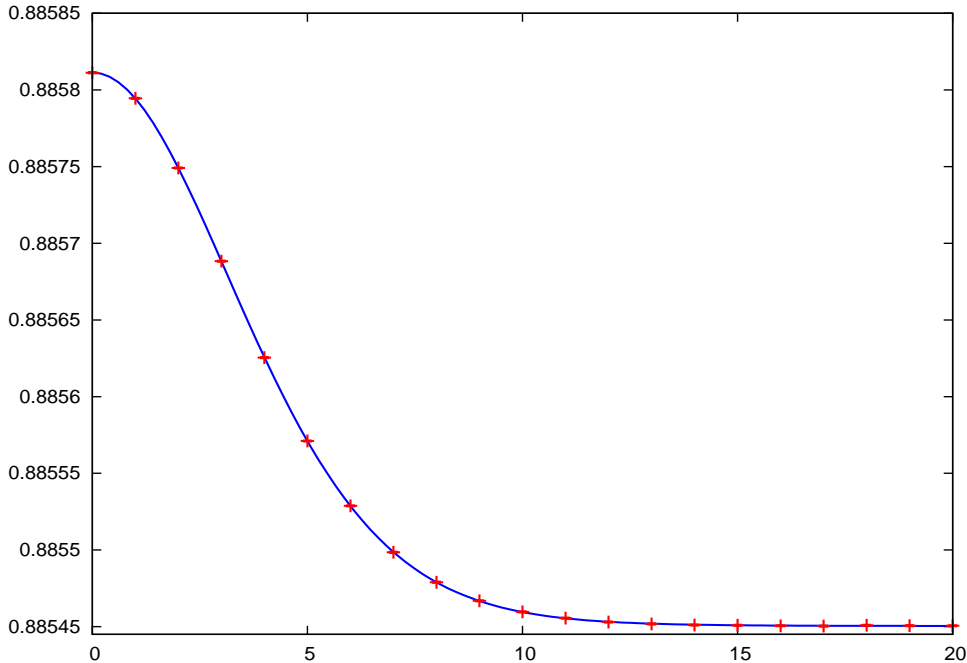}
\caption{The ratio $\frac{\langle \Phi_0 \Phi_r P_x \rangle}{\langle \Phi_0 \Phi_r \rangle}$
as a function of the transverse displacement  $x_3$ of the plaquette from the plane of the
string world-sheet. The distance between the two static sources is $r=17$. The solid
curve is a fit using eq.(\protect\ref{fitbell}).}
\label{bell}\end{figure}

In figure \ref{bell}, we show the bell-shaped correlation function 
${\langle \Phi_0 \Phi_r P_x \rangle}/{\langle \Phi_0 \Phi_r \rangle}$ as a function of the
transverse displacement of the plaquette from the plane of the string world-sheet. The data refer
to a pair of static sources at fixed distance $r=17$. The solid line is a fit of the
data using the ansatz (\ref{fitbell}).

The scale that determines the logarithmic broadening of the string at zero 
temperature was previously determined as $r_0 = 2.26(2)$ such that
$r_0 \sqrt{\sigma} = 0.364(3)$ \cite{Gli10}. It is important to note that $\sigma$ and
$r_0$ are the only low-energy parameters entering the 2-loop formula for the width of the
flux tube. Hence the theoretical prediction resulting from the low-energy effective
theory is completely determined without any additional adjustable parameter.

\begin{figure}[htb]
\centering
\includegraphics[width=0.84\textwidth]{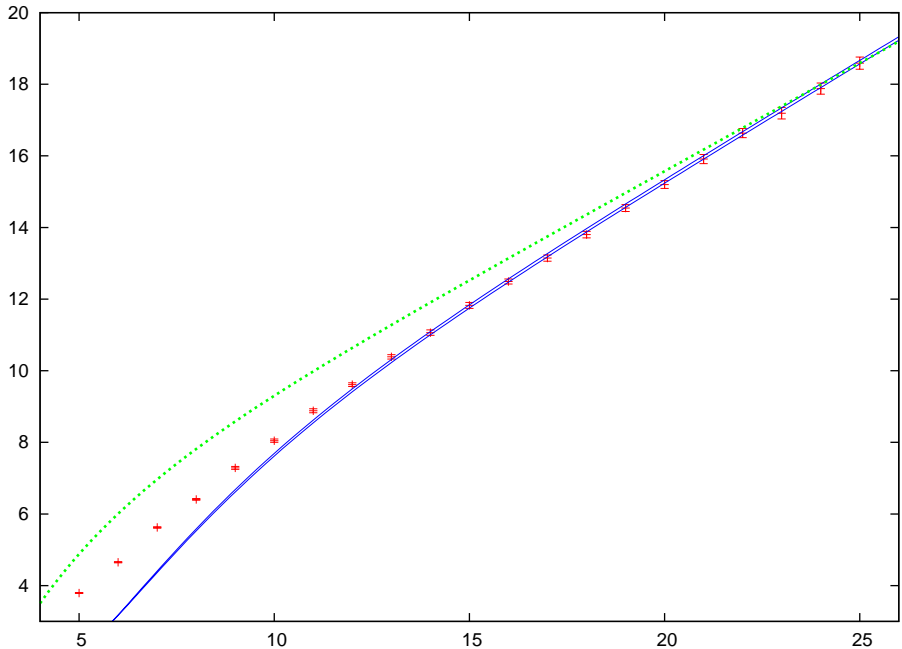}
\caption{The squared width of the confining string at its midpoint $w^2(r/2)$ as a
  function of the separation of the static sources $r$. The dotted line is the leading
  order prediction of the low-energy effective string theory using $r_0 = 2.26$. The two
  solid lines represent the 2-loop prediction using $r_0 = 2.26\pm 0.02$.}
\label{broadlinear}\end{figure}

In figure \ref{broadlinear} we report the observed dependence of the flux tube width on the
distance between the static sources as resulting from numerical Monte Carlo
simulations. The two solid lines represent the fully constrained prediction of the low-energy
effective string theory corresponding to the choices $r_0 = 2.26\pm 0.02$. 
The numerical data show the expected linear behavior for sufficiently large separations
of the sources and the agreement with the analytic formula turns out to be excellent.

Finally, it is important to notice that the next-to-leading expression (\ref{w2C}) for the
width is not sensitive to the next-to-leading correction to the string tension. In fact,
the 2-loop result for the width goes up to corrections of order $1/(\sigma \beta^2)$ and
so the corrections of order $1/(\sigma \beta^2)^2$ to the string tension will appear as
3-loop corrections to the width. Looking at the data for the width in figure
\ref{broadlinear}, the accuracy is such that we are not very sensitive to the
next-to-leading correction to the string tension and one cannot detect the tiny
difference that the direct measurement of the force in figure \ref{force} shows.

\section{Conclusions}

Using the L\"uscher-Weisz multi-level simulation technique, we have performed
high-accuracy numerical simulations of the confining string connecting two static
quarks in $(2+1)$-d $SU(2)$ Yang-Mills theory. The results for the finite
temperature string width have been compared with analytic
2-loop calculations performed in the corresponding systematic low-energy 
effective string 
theory. Since the low-energy parameters $\sigma$ and $r_0$ were already
determined at zero temperature, there were no further adjustable parameters in 
the comparison of the effective theory with the numerical simulations of the 
lattice Yang-Mills theory. Remarkably, there is perfect agreement between the 
underlying Yang-Mills theory and the effective string theory, thus confirming
the quantitative correctness of the latter. With sufficient numerical resources
(which are still modest on today's standards in lattice field theory) it is 
certainly feasible to extend our study to other gauge groups and to $(3+1)$
dimensions. A similar investigation in $(3+1)$-d $SU(3)$ Yang-Mills theory would
definitely be possible. In particular, it would be interesting to investigate 
whether the effective theory works equally well in the numerically accessible
regime as in the $(2+1)$-d case investigated here.

\section*{Acknowledgements}

We gratefully acknowledge helpful discussions with M.\ Caselle and F.\ Niedermayer.
This work is supported in part by funds provided by the Schweizerischer Nationalfonds
(SNF). The  ``Albert Einstein Center for Fundamental Physics'' at Bern University is 
supported by the ``Innovations- und Kooperationsprojekt  C-13'' of the Schweizerische 
Uni\-ver\-si\-t\"ats\-kon\-fe\-renz (SUK/CRUS).

\end{document}